\documentclass[twocolumn,aps,english,prl,nofootinbib,preprintnumbers]{revtex4}
\usepackage{mathrsfs}

\usepackage{mathrsfs}
\usepackage{graphicx}
\usepackage{amsmath, amssymb}
\usepackage{babel}
\usepackage{color}
\usepackage{slashed}
\usepackage{hyperref}
\usepackage[T1]{fontenc}
\usepackage[export]{adjustbox}

\begin{document}

\title{Model-independent determination of nuclear weak form factors and implications for Standard Model precision tests}

\author{Chien-Yeah Seng$^{1,2}$}

\affiliation{$^{1}$Facility for Rare Isotope Beams, Michigan State University, East Lansing, MI 48824, USA}
\affiliation{$^{2}$Department of Physics, University of Washington,
	Seattle, WA 98195-1560, USA}

\date{\today}

\begin{abstract}

We analyze the recoil corrections in superallowed beta decays of $T=1$, $J^P=0^+$ nuclei by fixing the mean square charged weak radius model-independently using the data of multiple charge radii across the nuclear isotriplet. By comparing to model estimations, we argue that the existing theory uncertainty in the statistical rate function $f$ might have been substantially underestimated. We discuss the implications of our proposed strategy for precision tests of Standard Model, including a potential alleviation of the first-row CKM unitarity deficit, and motivate new experiments for charge radii measurements. 

\end{abstract}

\maketitle

{\bf\boldmath Introduction} -- To confirm the recent observation of the first-row Cabibbo-Kobayashi-Maskawa (CKM) unitarity deficit~\cite{ParticleDataGroup:2022pth}, namely the apparent 3$\sigma$-deviation of the combination $|V_{ud}|^2+|V_{us}|^2+|V_{ub}|^2$ from the Standard Model (SM) prediction of 1, one needs to further improve the precision of $|V_{ud}|^2$ which weights the most in the unitarity sum. Currently, superallowed beta decays of $T=1$, $J^P=0^+$ nuclei report the most precise determination of $|V_{ud}|_{0^+}=0.97367(11)_\text{exp}(30)_\text{th}$~\cite{Hardy:2020qwl,Cirigliano:2022yyo}, but the precision from the free neutron decay is catching up, with $|V_{ud}|_n=0.97413(40)_\text{exp}(13)_\text{th}$~\cite{Cirigliano:2022yyo} if the single best measurement of the neutron lifetime~\cite{UCNt:2021pcg} and axial coupling~\cite{Markisch:2018ndu} are used (although the Particle Data Group (PDG) averages of these inputs leads to much a less precise $|V_{ud}|_n$ due to some internal tensions in the neutron dataset). Searching for small differences between the two determinations of $|V_{ud}|$ may open another window for the precision test of Standard Model (SM)~(see, e.g., \cite{Cirigliano:2012ab,Gonzalez-Alonso:2018omy,Crivellin:2020lzu,Crivellin:2021njn} and references therein), and for this purpose we must keep major sources of uncertainties in both avenues under control. For $|V_{ud}|_{0^+}$, this refers to the theory uncertainties which are the focus of this paper. 

Given the recent improvements of the nucleus-independent radiative corrections ~\cite{Seng:2018yzq,Seng:2018qru,Czarnecki:2019mwq,Seng:2020wjq,Hayen:2020cxh,Shiells:2020fqp,Gorchtein:2021fce,Cirigliano:2022hob}, the theory uncertainty in $|V_{ud}|_{0^+}$ comes mainly from nuclear structure effects. The quantity of interest is the so-called $\mathcal{F}t$-value, 
\begin{equation}
\mathcal{F}t=ft(1+\delta_\text{R}')(1+\delta_\text{NS}-\delta_\text{C})\propto |V_{ud}|_{0^+}^{-2}~,
\end{equation}
with $t$ the beta decay half-life. Most discussions of nuclear structure effects focus on $\delta_\text{NS}$ and $\delta_\text{C}$, which steam from the radiative corrections and the isospin symmetry breaking (ISB) correction respectively, and there are proposals to improve their precisions using ab-initio methods~\cite{Seng:2022cnq} and experiments~\cite{Seng:2022epj}. 
On the other hand, the ``outer'' radiative correction $\delta_\text{R}'$~\cite{Sirlin:1967zza,Sirlin:1987sy,Sirlin:1986cc} and the statistical rate function $f$~\cite{Towner:2014uta} were believed to be well under control; in particular, the combination $ft$ is often referred to as the ``experimental $ft$-value'' as if it was a pure experimental observable~\cite{Hardy:2014qxa,Hardy:2020qwl}. In this work we carefully examine the validity of this assertion.  

A large number of structure-dependent effects are included in the statistical rate function $f$ and were discussed extensively in literature (see, e.g. \cite{Hayen:2017pwg} and references therein). Most of them concern the interactions between the positron/atomic electrons and the nucleus of a finite size. Due to the electromagnetic nature, they can largely be fixed by the knowledge of the charge radius $R_\text{Ch}$ of the daughter nucleus, often very precisely measured, which guarantees the model-independence to certain extent. There is, however, one obvious exception, namely the recoil correction in the tree-level charged weak (CW) decay itself which scales as $q^2R_\text{CW}^2$, where $q^2$ is the squared momentum transfer and $R_\text{CW}^2$ is the mean square (MS) nuclear CW radius associated to the coupling of the nucleus to the $W$-boson; it influences the value of $f$ at (0.1-1)\% level. In general $R_\text{CW}$ and $R_\text{Ch}$ are quite different, so the information of a single nuclear charge radius cannot pin down this effect to a satisfactory level. To the best of our knowledge, all existing literature about $f$ handled the effects of nuclear CW form factors using simplified nuclear models~\cite{Wilkinson:1993hx,Hardy:2014qxa,Hayen:2017pwg}, which invalidate the claim of $ft$ being ``experimental''. Neither was the possible systematic error stemming from such modelings properly discussed.

Here we propose a simple method to restore the model-independence of this entry, namely to use two measured charge radii from the nuclear isotriplet to unambiguously determine $R_\text{CW}$, bearing small ISB corrections negligible in recoil effects. This idea was first pointed out by Holstein~\cite{Holstein:1974zf} but, for some reason, was not seriously implemented in subsequent analysis. We apply this strategy to 13 measured superallowed transitions and compare the outcomes with model-dependent estimates. We find that simplified models could result in systematic errors as large as $10^{-3}$ for medium and heavy nuclei. We discuss possible implications of this strategy for the precise $V_{ud}$ extraction and other precision tests of SM. 
We use them as new motivations for future experimental measurements of rare isotope charge radii.

{\bf\boldmath Tree-level decay rate}  -- 
In this work we concentrate on pure Quantum Chromodynamics (QCD) contributions to $f$, and neglect the electromagnetic interactions between the positron or atomic electrons and the nucleus. In this context, it is most convenient to start from a fully-relativistic description, where the kinematics of the superallowed nuclear beta decay $\phi_i(p_i)\rightarrow \phi_f(p_f)e^+(p_e)\nu_e(p_\nu)$ are exactly the same as the semileptonic kaon decay. In particular, all the analytic formula of the $K_{e3}$ decay rate in Ref.\cite{Seng:2021wcf} apply directly here upon simple replacements $V_{us}\rightarrow V_{ud}$, $K\rightarrow \phi_i$, $\pi\rightarrow \phi_f$. The tree-level decay amplitude reads:
\begin{equation}
\mathfrak{M}_0=-\frac{G}{\sqrt{2}}\bar{u}_\nu\gamma^\mu(1-\gamma_5)v_eF_\mu(p_f,p_i)~,
\end{equation}
where $G=G_FV_{ud}$, and $F_\mu$ is the nuclear matrix element of the charged weak current which defines two form factors:
\begin{eqnarray}
&&F_\mu(p_f,p_i)=\langle \phi(p_f)|J_\mu^{W\dagger}(0)|\phi_i(p_i)\rangle\nonumber\\
&&=f_+(q^2)(p_i+p_f)_\mu+f_-(q^2)(p_i-p_f)_\mu~,~\label{eq:CWFF}
\end{eqnarray}
with $q=p_i-p_f$. The contribution of $f_-(q^2)$ to the decay rate is simultaneously suppressed by kinematics ($m_e^2/M_i^2\ll 1$) and ISB, thus it is sufficient to retain only $f_+(q^2)$. We isolate the $q^2$-dependence of the latter by writing $f_+(q^2)=f_+(0)\bar{f}_+(q^2)$, where the leading $q^2$-dependence defines a MS CW radius $R_\text{CW}^2$:
\begin{equation}
\bar{f}_+(q^2)=1+(q^2/6)R_\text{CW}^2+\mathcal{O}(q^4)~.
\end{equation}
However, in practice one defines and scales out the so-called Fermi matrix element not at $q^2=0$ (which is inaccessible in beta decay processes) but at the static point $q^2=(M_i-M_f)^2$:
\begin{equation}
M_F\equiv f_+((M_i-M_f)^2)\approx f_+(0)(1+(M_i-M_f)^2R_\text{CW}^2/6)~,
\end{equation} 
in contrast to the usual treatment in kaon decay that scales out $f_+(0)$. Notice that $M_F\rightarrow M_F^0=\sqrt{2}$ in the absence of ISB.

After summing up the lepton spins, the squared amplitude reads:
\begin{equation}
|\mathfrak{M}_0|^2=G^2f_+^2(q^2)H(+1,+1)~,
\end{equation}
where 
\begin{eqnarray}
H(+1,+1)&=&2M_i^4[4(1-y)(y+z-1)-4r_f\nonumber\\
&&+r_e(r_f+4y+3z-3)-r_e^2]
\end{eqnarray}
depends only on two scalar variables, $y=2p_i\cdot p_e/M_i^2$ and $z=2p_i\cdot p_f/M_i^2$ (with $r_e\equiv m_e^2/M_i^2$ and $r_f^2\equiv M_f^2/M_i^2$). Integrating out $z$ and taking $M_{i,f}\rightarrow \infty$ give the following differential decay rate in the parent nucleus's rest frame:
\begin{equation}
\frac{d\Gamma}{dE_e}\approx \frac{G^2}{2\pi^3}|M_F|^2F(Z_f,E_e)|\vec{p}_e|E_e(E_m-E_e)^2S\label{eq:rate}
\end{equation}
where $E_m\equiv (M_i^2-M_f^2+m_e^2)/(2M_i)\approx M_i-M_f$ is the positron's end-point energy, and we add the Fermi function $F(Z_f,E_e)$~\cite{Wilkinson:1970cdv} manually just to improve the numerical accuracy. The quantity $S$ denotes the pure-QCD contribution to the shape factor:
\begin{equation}
S=1+\frac{R_\text{CW}^2m_e^2}{9}\left(1-\frac{3E_m^2}{m_e^2}+\frac{4E_m E_e}{m_e^2}+\frac{2E_m}{E_e}-\frac{4E_e^2}{m_e^2}\right)
\end{equation}
in agreement with existing literature~\cite{Behrens1982ElectronRW,Wilkinson:1993hx,Hayen:2017pwg} (upon taking $1-3E_m^2/m_e^2\approx -3E_m^2/m_e^2$).

{\bf\boldmath Isospin relation between CW and charge radii}  -- In his famous review~\cite{Holstein:1974zf}, Holstein derived the relation between the vector CW form factors and the difference between the parent and daughter nucleus's electromagnetic form factors using the conserved vector current (CVC) condition. Here we present a generalized version, not restricted to the parent-daughter pair but any two nuclei within the isotriplet, in order to better make use of the currently-available data of charge radii.

First, using the spatial translational symmetry and working in a modified Breit frame ($E_i=E_f$), the $\mu=0$ component of Eq.\eqref{eq:CWFF} gives: 
\begin{eqnarray}
&&f_+(q^2)=\int d^3x e^{-i\vec{q}\cdot\vec{x}}\langle \phi_f|\rho_W(r)|\phi_i\rangle\nonumber\\
&&=\int d^3x\left(1-\frac{(\vec{q}\cdot\vec{x})^2}{2}+...\right)\langle \phi_f|\rho_W(r)|\phi_i\rangle
\end{eqnarray}
where $q^2=-\vec{q}^2$, $\rho_W(r)$ is the time component of the charged weak current operator which depends only on $r=|\vec{x}|$ due to rotational symmetry, and $|\phi_{i,f}\rangle$ are quantum mechanical external nuclear states that normalize to 1. Our interest is in the second term which, after symmetric replacement, provides a formal definition of $R_\text{CW}^2$: 
\begin{equation}
R_\text{CW}^2=\frac{1}{f_+(0)}\langle \phi_f|\int d^3x r^2\rho_W(r)|\phi_i\rangle~.\label{eq:RCWformal}
\end{equation}
The r.h.s may be expressed in terms of the isovector monopole operator:
\begin{equation}
\vec{M}^{(1)}\equiv\int d^3xr^2\psi^\dagger(x)\frac{\vec{\tau}}{2}\psi(x)~,
\end{equation}
where $\psi=(d,u)^\text{T}$ is the light quark doublet field (here we adopt the nuclear theory convention of isospin, $T_{z,u}=-1/2$). Rank-1 irreducible tensors in the isospin space can be constructed as: $M_0^{(1)}=M_z^{(1)}$, $M_{\pm 1}^{(1)}=\mp (M_x^{(1)}\pm i M_y^{(1)})/\sqrt{2}$. Finally, we may take $f_+(0)\rightarrow\sqrt{2}$ in Eq.\eqref{eq:RCWformal} because ISB on top of a recoil correction is negligible. Therefore we obtain:
\begin{equation}
R_\text{CW}^2=-\langle \phi_f|M_{+1}^{(1)}|\phi_i\rangle~.\label{eq:RCWsimple}
\end{equation}

We want to relate $R_\text{CW}^2$ to the MS charge radius of a nucleus $\phi$, defined as:
\begin{eqnarray}
&&R_{\text{Ch},\phi}^2=\frac{1}{Z_\phi}\langle \phi|\int d^3x r^2 \rho_\text{Ch}(r)|\phi\rangle\nonumber\\
&&\!\!\!\!\!\!\!\!\!\!\!=\frac{1}{Z_\phi}\langle\phi|\int d^3x r^2\left(\frac{1}{6}\psi^\dagger \psi-\frac{1}{3}s^\dagger s-\psi^\dagger\frac{\tau^3}{2}\psi\right)|\phi\rangle~,\label{eq:Rp2}
\end{eqnarray}
with $Z_\phi$ the atomic number of $\phi$. For simplicity, we will label $Z,R_\text{Ch}$ of an isotriplet nuclear state $|1,T_z\rangle$ as $Z_{T_z}$, $R_{\text{Ch},T_z}$ respectively.
The r.h.s of the second line in Eq.\eqref{eq:Rp2} consists of two isoscalar terms and an isovector term; the last is just the nuclear matrix element of $M_0^{(1)}$. By constructing the difference between $Z_\phi R_{\text{Ch},\phi}^2$ of two nuclei within the same isotriplet, the isosinglet pieces drop out and the remaining isovector term can then be related to Eq.\eqref{eq:RCWsimple} in the isospin-symmetric limit through the Wigner-Eckart theorem:
\begin{equation}
\langle 1,T_{zb}|M^{(1)}_{m}|1,T_{za}\rangle=C_{1,T_{za};1,m}^{1,1;1,T_{zb}}\langle 1||M^{(1)}||1\rangle~,
\end{equation}
with $C_{1,T_{za};1,m}^{1,1;1,T_{zb}}$ the Clebsch-Gordan coefficient and $\langle 1||M^{(1)}||1\rangle$ the reduced matrix element. With this we finally obtain:
\begin{eqnarray}
R_\text{CW}^2&=&R_{\text{Ch},1}^2+Z_0(R_{\text{Ch},0}^2-R_{\text{Ch},1}^2)\nonumber\\
&=&R_{\text{Ch},1}^2+\frac{Z_{-1}}{2}(R_{\text{Ch},-1}^2-R_{\text{Ch},1}^2)~,\label{eq:isospin}
\end{eqnarray}
where we have used $Z_1=Z_0-1=Z_{-1}-2$. 

\begin{table*}[t]
	\begin{centering}
	\begin{tabular}{|c|c|c|c|c|c|}
		\hline 
		$A$ & $R_{\text{Ch},-1}$ (fm) & $R_{\text{Ch},0}$ (fm) & $R_{\text{Ch},1}$ (fm) & $R_{\text{Ch,1}}^{2}$ (fm$^{2}$) & $R_{\text{CW}}^{2}$ (fm$^{2}$)\tabularnewline
		\hline 
		\hline 
		10 & $_{6}^{10}$C & $_{5}^{10}$B(ex) & $_{4}^{10}$Be: 2.3550(170)$^{a}$ & 5.546(80) & N/A\tabularnewline
		\hline 
		14 & $_{8}^{14}$O & $_{7}^{14}$N(ex) & $_{6}^{14}$C: 2.5025(87)$^{a}$ & 6.263(44) & N/A\tabularnewline
		\hline 
		18 & $_{10}^{18}$Ne: 2.9714(76)$^{a}$ & $_{9}^{18}$F(ex) & $_{8}^{18}$O: 2.7726(56)$^{a}$ & 7.687(31) & 13.40(53)\tabularnewline
		\hline 
		22 & $_{12}^{22}$Mg: 3.0691(89)$^{b}$ & $_{11}^{22}$Na(ex) & $_{10}^{22}$Ne: 2.9525(40)$^{a}$ & 8.717(24) & 12.93(71)\tabularnewline
		\hline 
		26 & $_{14}^{26}$Si & $_{13}^{26m}$Al & $_{12}^{26}$Mg: 3.0337(18)$^{a}$ & 9.203(11) & N/A\tabularnewline
		\hline 
		30 & $_{16}^{30}$S & $_{15}^{30}$P(ex) & $_{14}^{30}$Si: 3.1336(40)$^{a}$ & 9.819(25) & N/A\tabularnewline
		\hline 
		34 & $_{18}^{34}$Ar: 3.3654(40)$^{a}$ & $_{17}^{34}$Cl & $_{16}^{34}$S: 3.2847(21)$^{a}$ & 10.789(14) & 15.62(54)\tabularnewline
		\hline 
		38 & $_{20}^{38}$Ca: 3.467(1)$^{c}$ & $_{19}^{38m}$K: 3.437(4)$^{d}$ & $_{18}^{38}$Ar: 3.4028(19)$^{a}$ & 11.579(13) & 15.99(28)\tabularnewline
		\hline 
		42 & $_{22}^{42}$Ti & $_{21}^{42}$Sc: 3.5702(238)$^{a}$ & $_{20}^{42}$Ca: 3.5081(21)$^{a}$ & 12.307(15) & 21.5(3.6)\tabularnewline
		\hline 
		46 & $_{24}^{46}$Cr & $_{23}^{46}$V & $_{22}^{46}$Ti: 3.6070(22)$^{a}$ & 13.010(16) & N/A\tabularnewline
		\hline 
		50 & $_{26}^{50}$Fe & $_{25}^{50}$Mn: 3.7120(196)$^{a}$ & $_{24}^{50}$Cr: 3.6588(65)$^{a}$ & 13.387(48) & 23.2(3.8)\tabularnewline
		\hline 
		54 & $_{28}^{54}$Ni: 3.738(4)$^{e}$ & $_{27}^{54}$Co & $_{26}^{54}$Fe: 3.6933(19)$^{a}$ & 13.640(14) & 18.29(92)\tabularnewline
		\hline 
		62 & $_{32}^{62}$Ge & $_{31}^{62}$Ga & $_{30}^{62}$Zn: 3.9031(69)$^{b}$ & 15.234(54) & N/A\tabularnewline
		\hline 
		66 & $_{34}^{66}$Se & $_{33}^{66}$As & $_{32}^{66}$Ge & N/A & N/A\tabularnewline
		\hline 
		70 & $_{36}^{70}$Kr & $_{35}^{70}$Br & $_{34}^{70}$Se & N/A & N/A\tabularnewline
		\hline 
		74 & $_{38}^{74}$Sr & $_{37}^{74}$Rb: 4.1935(172)$^{b}$ & $_{36}^{74}$Kr: 4.1870(41)$^{a}$ & 17.531(34) & 19.5(5.5)\tabularnewline
		\hline 
	\end{tabular}
		\par\end{centering}
	\caption{\label{tab:RCW}Determinations of $R_\text{CW}^2$ based on available data of nuclear charge radii for isotriplets in measured superallowed decays. Superscripts denote the source of data: Ref.\cite{Angeli:2013epw}$^a$, Ref.\cite{Li:2021fmk}$^b$, Ref.\cite{C38chargeradius}$^c$, Ref.\cite{Bissell:2014vva}$^d$ and Ref.\cite{Pineda:2021shy}$^e$.}
\end{table*}

Eq.\eqref{eq:isospin} is the central result of this work: it says that $R_\text{CW}^2$ can be determined model-independently, modulo negligible ISB corrections, if the charge radius of at least two nuclei within the isotriplet are known experimentally. There are two terms at the r.h.s of Eq.\eqref{eq:isospin}; the first term is the MS charge radius of the most stable $T_z=+1$ nucleus, while the second term involves a difference $R_{\text{Ch},a}^2-R_{\text{Ch},b}^2$. Nevertheless, this term is numerically comparable to the first term because it is multiplied to a large factor $Z$; in fact, it is also the main source of error because the experimental uncertainties in $R_\text{Ch}^2$ are enhanced by the same factor. Therefore, we expect the error of $R_\text{CW}^2$ determined with this method to be roughly an order of magnitude larger than that of the individual $R_\text{Ch}^2$.

We present our model-independent determination of $R_\text{CW}^2$ in Table~\ref{tab:RCW} based on the currently-available data of charge radii for nuclear isotriplets involved in measured superallowed transitions~\cite{Angeli:2013epw,Li:2021fmk,C38chargeradius,Bissell:2014vva,Pineda:2021shy}. One observes that in many cases it is substantially larger than $R_\text{Ch}^2$, which signifies the importance of the ``difference'' term in Eq.\eqref{eq:isospin}. Also, unlike the charge radius, $R_\text{CW}$ does not seem to increase monotonically with the mass number $A$, which makes an accurate theory modeling of its value much more difficult.

{\bf\boldmath Recoil effects: Experiment vs model} -- Despite being known since the 1970s, we are not aware of any literature that seriously implemented the aforementioned idea in their numerical analysis of $f$; instead, most of them resort to nuclear models. For instance, Hardy and Towner~\cite{Hardy:2004id} computed the nuclear form factors directly using the impulse approximation, where nucleons in a nucleus are treated as non-interacting, and the nuclear matrix element of a one-body operator $\hat{O}$ is expressed as a product of the single-nucleon matrix element of $\hat{O}$ (with the $q^2$-dependence neglected) and the one-body density matrix element, the latter is computed with shell model. To what extent such an approximation captures the correct $q^2$-dependence of the nuclear form factors is far from transparent. 
A more traceable method was introduced by Wilkinson~\cite{Wilkinson:1993hx}, who estimated the difference between $R_\text{CW}^2$ and $R_\text{Ch}^2$ using shell model and a modified-Gaussian charge distribution:
\begin{equation}
R_\text{CW}^2-R_\text{Ch}^2\approx \frac{4}{3(5A'+2)}\frac{4n+2l-1}{5}R_\text{Ch}^2~,\label{eq:RCWmodel}
\end{equation}   
where $\{n,l\}$ are the shell-model quantum numbers of the single active nucleon that undergoes the beta decay, and $A'$ is a parameter of the modified-Gaussian charge distribution fixed by the condition $2/(2+3A')=Z_{l=0}/Z$ for the parent nucleus.
As we will see later that the effects of $S$ to the total decay rate can reach 0.1\% or above for medium and heavy nuclei, theory errors in the $R_\text{CW}$-modeling could lead to corrections at (0.01-0.1)\% level which are relevant for the precise extraction of $V_{ud}$.

\begin{table}
	\begin{centering}
		\begin{tabular}{|c|c|c|c|c|}
			\hline 
			Parent & $\frac{\Gamma_{\text{exp}}-\Gamma_{0}}{\Gamma_{\exp}}$ & $\frac{\Gamma_{\text{exp}}-\Gamma_{\text{mod}}^{0}}{\Gamma_{\exp}}$ & $\frac{\Gamma_{\text{exp}}-\Gamma_{\text{mod}}}{\Gamma_{\exp}}$ & $\frac{\delta f}{f}$ in \cite{Hardy:2020qwl}\tabularnewline
			\hline 
			\hline 
			$T_{z,i}=-1$ &  &  &  & \tabularnewline
			\hline 
			$^{18}$Ne & -0.06(0) & -0.03(0) & -0.02(0) & 0.13\tabularnewline
			\hline 
			$^{22}$Mg & -0.10(1) & -0.03(1) & -0.03(1) & 0.03\tabularnewline
			\hline 
			$^{34}$Ar & -0.29(1) & -0.09(1) & -0.06(1) & 0.01\tabularnewline
			\hline 
			$^{38}$Ca & -0.36(1) & -0.10(1) & -0.07(1) & 0.01\tabularnewline
			\hline 
			$^{42}$Ti & -0.55(9) & -0.23(9) & -0.19(9) & 0.02\tabularnewline
			\hline 
			$^{50}$Fe & -0.82(13) & -0.35(13) & -0.29(13) & 0.40\tabularnewline
			\hline 
			$^{54}$Ni & -0.75(4) & -0.19(4) & -0.13(4) & 0.27\tabularnewline
			\hline 
			$T_{z,i}=0$ &  &  &  & \tabularnewline
			\hline 
			$^{34}$Cl & -0.23(1) & -0.07(1) & -0.05(1) & 0.00\tabularnewline
			\hline 
			$^{38m}$K & -0.29(1) & -0.08(1) & -0.05(1) & 0.00\tabularnewline
			\hline 
			$^{42}$Sc & -0.45(8) & -0.19(8) & -0.15(8) & 0.01\tabularnewline
			\hline 
			$^{50}$Mn & -0.71(12) & -0.30(12) & -0.25(12) & 0.00\tabularnewline
			\hline 
			$^{54}$Co & -0.66(3) & -0.17(3) & -0.11(3) & 0.02\tabularnewline
			\hline 
			$^{74}$Rb & -1.17(33) & -0.12(33) & -0.03(33) & 0.20\tabularnewline
			\hline 
		\end{tabular}
		\par\end{centering}
	\caption{\label{tab:result} Comparison between different determinations of the superallowed decay rate. The uncertainty comes primarily from $R_\text{CW}$ in $\Gamma_\text{exp}$. All number are in \%.}
\end{table}

Based on the data in Table~\ref{tab:RCW}, we can immediately study the effect of $S$ to the total decay rate model-independently for 13 out of 23~\cite{Hardy:2020qwl} measured superallowed transitions. We integrate $E_e$ in Eq.\eqref{eq:rate} to obtain a total decay rate $\Gamma$, and we do it in four different ways: (1) $\Gamma_\text{exp}$ denotes our model-independent determination making use of the experimental values of $R_\text{CW}$ given in Table~\ref{tab:RCW}; 
(2) Denoted by $\Gamma_0$, we take $S=1$, i.e. completely neglect the recoil correction; (3) Denoted by $\Gamma_\text{mod}^0$, we replace $R_\text{CW}$ in $S$ by the charge radius of the most stable $T_z=+1$ isotope $R_\text{Ch,1}$; (4) Denoted by $\Gamma_\text{mod}$, we substitute $R_\text{CW}^2$ by Wilkinson's shell-model estimate, Eq.\eqref{eq:RCWmodel}. What we are interested is the relative difference between the experimental result and the modelings (2)--(4), so we use the ratio $(\Gamma_\text{exp}-\Gamma_i)/\Gamma_\text{exp}$ to represent the systematic error induced by the modeling type $i$.

Our results are summarized in Table~\ref{tab:result}. From the first column we see the size of the recoil correction: it is negative and at (0.1-1)\% level as we advertised before, and increases with the mass number. The second column shows the induced systematic error if one would na\"{i}vely replace $R_\text{CW}$ by $R_\text{Ch}$; we find that it ranges from -0.03\% to -0.35\%, indicating again the significance of the ``difference'' term in Eq.\eqref{eq:isospin}. The third column shows how the modeling of $R_\text{CW}$ in Eq.\eqref{eq:RCWmodel} saves the situation, and we find that in most cases it only very mildly improves the accuracy, indicating that Eq.\eqref{eq:RCWmodel} still largely underestimates the difference $R_\text{CW}^2-R_\text{Ch}^2$. Finally, in the fourth column we show the quoted relative uncertainty of the statistical rate function $f$ in the most recent review by Hardy and Towner, Ref.\cite{Hardy:2020qwl}. We find that, in most cases the central values in the third column largely exceed the numbers in the fourth column. Of course the comparison is not totally fair because it is not clear at this point that the method used in Ref.\cite{Hardy:2020qwl} to effectively handle $R_\text{CW}$ is similar to that in Eq.\eqref{eq:RCWmodel}. Nevertheless, it still provides a strong indication that the systematic error in $f$ due to theory modelings of the CW form factor might have been underestimated. 

{\bf\boldmath Final discussions} -- To fully make use of our model-independent determination of $R_\text{CW}$, one should carefully sort out the theory modelings of the nuclear CW form factors in recent literature that compute $f$, e.g. Refs.\cite{Hardy:2004id,Hardy:2008gy,Hardy:2014qxa,Hardy:2020qwl,Hayen:2017pwg}, and replace them consistently by the experimental results. Also, to incorporate the Coulomb effects between the positron and the nucleus, updated charge distributions that are fully compatible with the most recent charge radii measurements are needed. This is not restricted to the 13 transitions that we analyzed above, but should also be applied to all remaining superallowed transitions once new data of charge radii are available in the future. Also, a straightforward generalization of Eq.\eqref{eq:isospin} to $T=1/2$ systems provides model-independent determination of charged weak form factors of neutron and nuclear mirrors~\cite{NaviliatCuncic:2008xt}, both serving as alternative avenues to measure $|V_{ud}|$. 

Our finding hints towards a possible solution of the CKM anomaly. For instance, if we would na\"{\i}vely reduce the overall $\mathcal{F}t$-value by the average of the central values of column 3 in Table~\ref{tab:result}, i.e. 0.11\%, then the central value of $|V_{ud}|_{0^+}$ would increase from 0.97367 to 0.97421, almost recovering the pre-2018 value of 0.97417(21)~\cite{Hardy:2014qxa} and largely restoring the first-row CKM unitarity. A more robust number, of course, has to come from a combination of experimental data and a comprehensive re-analysis of all existing models as described above, which we save for a future work.
Recall also that the alignment of the $\mathcal{F}t$-values across different nuclei is used to test the CVC hypothesis, to constrain scalar currents and to test the reliability of nuclear model calculations of the ISB correction $\delta_\text{C}$~\cite{Towner:2010bx}. Therefore, possible nucleus-dependent alterations of the $f$-values could lead to modified interpretations of these constraints. 
Besides, the experimental determination of $R_\text{CW}$ also improves the theory handle of other CW processes that involve the same form factor, for example the neutrino-nucleus scattering $\nu\phi\rightarrow \ell^+\phi'$ where the momentum exchange is much larger and the effects of the form factors are more significant.


As indicated in Table~\ref{tab:RCW}, for nuclear isotriplets with $A$=10, 14, 26, 30, 46 and 62, the addition of one single charge radius measurement will already provide sufficient input for the model-independent $R_\text{CW}$ determination, and with them we can sharpen our theory prediction of $f$ for 8 more superallowed transitions, further improving the reliability of the $|V_{ud}|_{0^+}$-extraction. These measurements may be performed, for example, at the BECOLA facility at FRIB~\cite{MINAMISONO201385}, or within the context of the muX experiment at PSI~\cite{Knecht:2020npz}. We hope the discussions above provide convincing new motivations for the planning of future experimental programs for charge radii measurements of rare isotopes at these facilities.

\begin{acknowledgments}
	
{\bf\boldmath Acknowledgments} -- The author thanks Ayala Glick-Magid, Mikhail Gorchtein, Leendert Hayen and Gerald Miller for useful conversations, and is grateful to Kei Minamisono for providing references for various charge radii. This work is supported in
part by the U.S. Department of Energy (DOE), Office of Science, Office of Nuclear Physics, under the FRIB Theory Alliance award DE-SC0013617, and by the DOE grant DE-FG02-97ER41014.
	
\end{acknowledgments}

\bibliography{RCW_ref}

\end{document}